\documentclass[aps,reprint,a4paper,nobalancelastpage]{revtex4-1}
\usepackage{amssymb,amsmath}
\usepackage{subfigure}
\usepackage{graphicx}
\usepackage{mathptmx}
\usepackage{enumerate}
\usepackage{CJK}
\usepackage{color}
\usepackage[breaklinks=true,colorlinks=true,bookmarks=false,urlcolor=blue,
citecolor=blue,linkcolor=blue,hypertexnames=false]{hyperref}
\usepackage{fancyhdr}
\usepackage{lipsum}

\def\unit#1{\mathord{\thinspace\rm #1}}
\def\func#1{\mathop{\rm #1}\nolimits}

\definecolor{ur}{rgb}{0.071,0.388,0.678}

\fancypagestyle{plain}{
\fancyhf{}
\fancyfoot[C]{\small \thepage}
}

\begin{document}

\begin{CJK*}{UTF8}{}

%\title{Gate-Controllable One-Dimensional Superlattice in Graphene}
\title{Cloning of Zero Modes in One-Dimensional Graphene Superlattices}

\author{Wun-Hao Kang (\CJKfamily{bsmi}{康文豪})}
\affiliation{Department of Physics, National Cheng Kung University, Tainan 70101, Taiwan}

\author{Szu-Chao Chen (\CJKfamily{bsmi}{陳思超})}
\affiliation{Department of Physics, National Cheng Kung University, Tainan 70101, Taiwan}

\author{Ming-Hao Liu (\CJKfamily{bsmi}{劉明豪})}
\email{minghao.liu@phys.ncku.edu.tw}
\affiliation{Department of Physics, National Cheng Kung University, Tainan 70101, Taiwan}

\date{\today}

\begin{abstract}

One-dimensional (1D) graphene superlattices have been predicted to exhibit zero-energy modes a decade ago, but an experimental proof has remained missing. Motivated by a recent experiment \cite{Note1} that could possibly shed light on this, here we perform quantum transport simulations for 1D graphene superlattices, considering electrostatically simulated potential profiles as realistic as possible. Combined with the analysis on the corresponding miniband structures, we find that the zero modes generated by the 1D superlattice potential can be further cloned to higher energies, which are also accessible by tuning the average density. Our multiterminal transverse magnetic focusing simulations further reveal the modulation-controllable ballistic miniband transport for 1D graphene superlattices. A simple idea for creating a perfectly symmetric periodic potential with strong modulation is proposed at the end of this work, generating well aligned zero modes up to 6 within a reasonable gate strength.

\end{abstract}


\maketitle

\end{CJK*}

\thispagestyle{plain}

Electric potential varying periodically in space in a length scale much longer than the lattice constant of the host lattice on which the potential is applied forms the so-called superlattice. In the case of graphene, superlattice potential has been shown to induce extra Dirac points and group velocity renormalization \cite{Park2008b}. Among various types of graphene superlattices, one-dimensional (1D) periodic potential is shown to give rise to velocity renormalization in the original Dirac cone and the emergence of additional zero-energy Dirac points \cite{Park2009a,Brey2009}. As a result, 1D graphene superlattices are predicted to exhibit anisotropic transport \cite{Barbier2010,Burset2011,Wu2012}, Hall conductivity step of $4(2N+1)e^2/h$ with $N=0,1,2,\cdots$ \cite{Park2008b} and conductance resonances accompanying the generation of new zero modes \cite{Brey2009}. 

A decade has passed. Experimental progresses on 1D graphene superlattices remain rather limited \cite{Dubey2013,Drienovsky2014}. Motivated by the recent technical breakthrough on creating gate-controllable graphene superlattices \cite{Forsythe2018,Drienovsky2018} and particularly a very recent work \footnote{Dean group; in preparation and possibly submitted.} on 1D graphene superlattices using periodically patterned SiO$_2$ substrate, here we re-examine this relatively old issue based on our theoretical tool kits of tight-binding quantum transport simulations and finite-element-based electrostatic simulations using \textsc{FEniCS} \cite{fenics} and \textsc{Gmsh} \cite{gmsh}, along with calculations of continuum-model-based superlattice miniband structures and their corresponding density of states. We first model the device of \cite{Note1} and simulate its transport properties, and then move on to an optimized design for exploring ballistic miniband transport properties. At the end of this paper, we discuss how to achieve a perfectly symmetric superlattice potential, generating zero-energy modes up to 6 which are further cloned to higher energy minibands accessible to transport by tuning the average density.

\begin{figure}[b]
\includegraphics[width=\columnwidth]{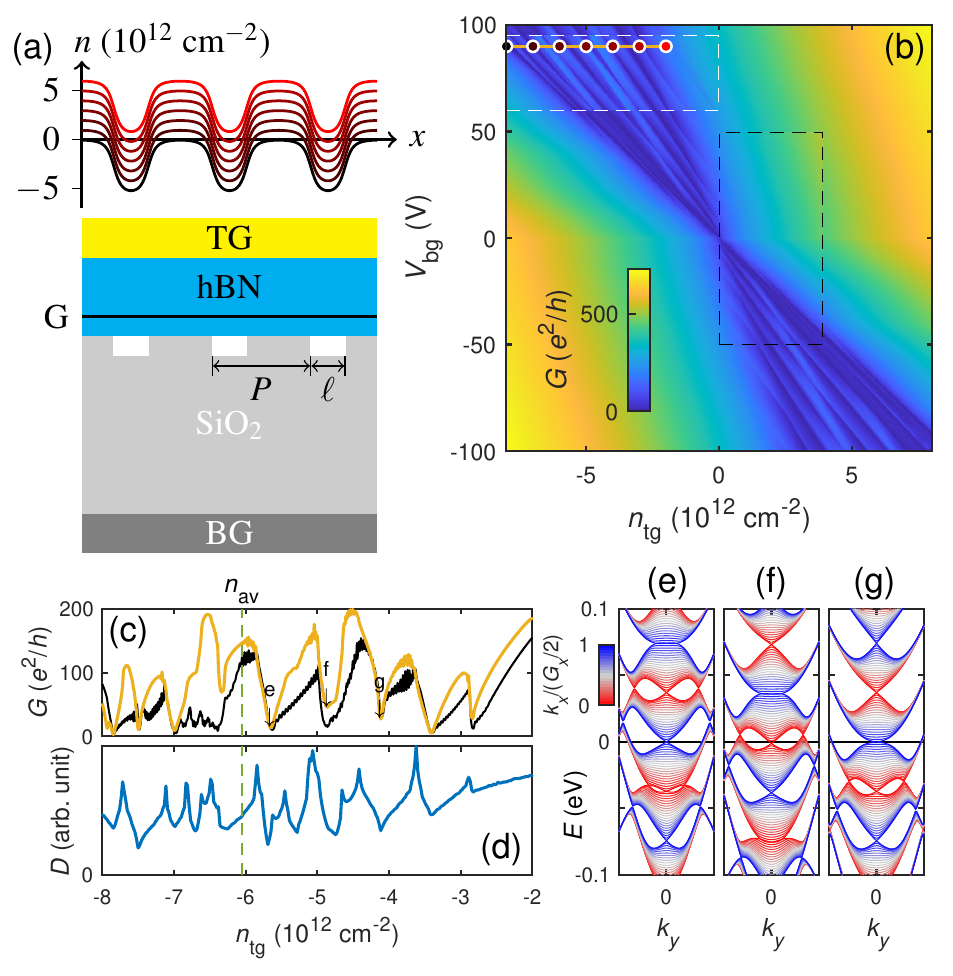}
\caption{Modeling and simulations for the device of \cite{Note1}. (a) Schematic for the modeled device. Exemplary carrier density profiles based on the electrostatic simulation with $P=55\unit{nm}$ and $\ell=20\unit{nm}$, considering gate voltages marked on (b), which reports a two-terminal conductance from transport simulation. Black and white dashed boxes mark regions where different data are overlayed as explained in the main text. (c) Conductance line cuts corresponding to the horizontal orange line on (b); black and orange lines are from the main and white-dash-boxed maps of (b), respectively. (d) Fermi density of states $D(E=0)$ as a function of the top gate density $n_\mathrm{tg}$, showing the same range as (c). (e)--(g) Miniband structures corresponding to gating configurations marked on (c).}
\label{fig1}
\end{figure}

Figure \ref{fig1}(a) shows a side-view schematic of the gating geometry used in \cite{Note1}, where the periodicity $P=55\unit{nm}$ and the etch length $15\unit{nm}\leq\ell\leq 20\unit{nm}$ were designed. Due to the periodic modulation of the SiO$_2$ substrate, the back gate capacitance is position-dependent, $C_b=C_b(x)$, and the top gate separated from the graphene sample by a hexagonal boron nitride (hBN) layer of dielectrics can be assumed to contribute a uniform carrier density, $n_\mathrm{tg} = (C_t/e)V_\mathrm{tg}$, where $C_t$ is the uniform top gate capacitance and $V_\mathrm{tg}$ is top gate voltage. The net carrier density is given by $n(x)=n_\mathrm{tg}+[C_b(x)/e]V_\mathrm{bg}$, and the corresponding superlattice potential given by $V(x) = -\hbar v_F\sqrt{\pi|n(x)|}\func{sgn}[n(x)]$ is to be added to the tight-binding model Hamiltonian as the onsite energy term. Taking $\ell=20\unit{nm}$, a few exemplary carrier density profiles are shown in the upper part of Fig.\ \ref{fig1}(a) \footnote{See Supplementary Material at \url{xxxx} for further information about electrostatic simulations, two-point conductance simulation, and a brief proof of symmetric periodic potential using two sets of periodic gates.}. Various colors of the carrier densities $n(x)$ shown in Fig.\ \ref{fig1}(a) correspond to the dots marked with respective colors on Fig.\ \ref{fig1}(b), which reports the two-terminal conductance through a graphene sample subject to a back gate dielectric etched with 21 trenches, as a function of $n_\mathrm{tg}$ and $V_\mathrm{bg}$. With the minimal tight-binding model for bulk graphene ($N_z=2$ zigzag ribbons with periodic boundary hopping modulated by the Bloch phase \cite{Wimmer2008,Liu2012a}), we first calculate the normalized conductance $g = \int_{-k_F}^{k_F} T(E_F;k_y) dk_y$, and the ballistic conductance \cite{Beenakker2008} is then obtained by $G=(W/3\pi a)g$, where $a$ is the honeycomb lattice spacing, either pristine or scaled \cite{Liu2015}, and $W$ is the assumed sample width. We have also performed transport calculations using a finite-width armchair ribbon scaled by a factor of $s_f = 10$ \cite{Liu2015}. Part of the result is overlayed on Fig.\ \ref{fig1}(b) marked by the black dashed box. Considering $W=1000\unit{nm}$, the consistency is extremely satisfactory \cite{Note2}.

At a glance, the hourglass-like conductance map of Fig.\ \ref{fig1}(b) may appear similar to those of graphene devices with multiple \textit{pn} junctions, exhibiting complex Fabry-P\'erot interference patterns \cite{Dubey2013,Drienovsky2014}. However, a closer inspection reveals that the significant drop of the conductance approaching to zero cannot be explained by simple pictures of destructive wave interference. A typical line cut of the conductance as a function of $n_\mathrm{tg}$ showing non-sinusoidal dips is shown by the black curve in Figure \ref{fig1}(c), corresponding to the horizontal orange line marked on (b). Due to the considered geometry close to the experiment, however, the resulting superlattice potential profiles are not perfectly periodic. To closely compare with superlattice miniband structures, in the following we consider perfectly periodic back gate capacitance obtained by numerically repeating the central period. For the subtle comparison between the two models, see \cite{Note2}. 

\begin{figure}
\includegraphics[width=\columnwidth]{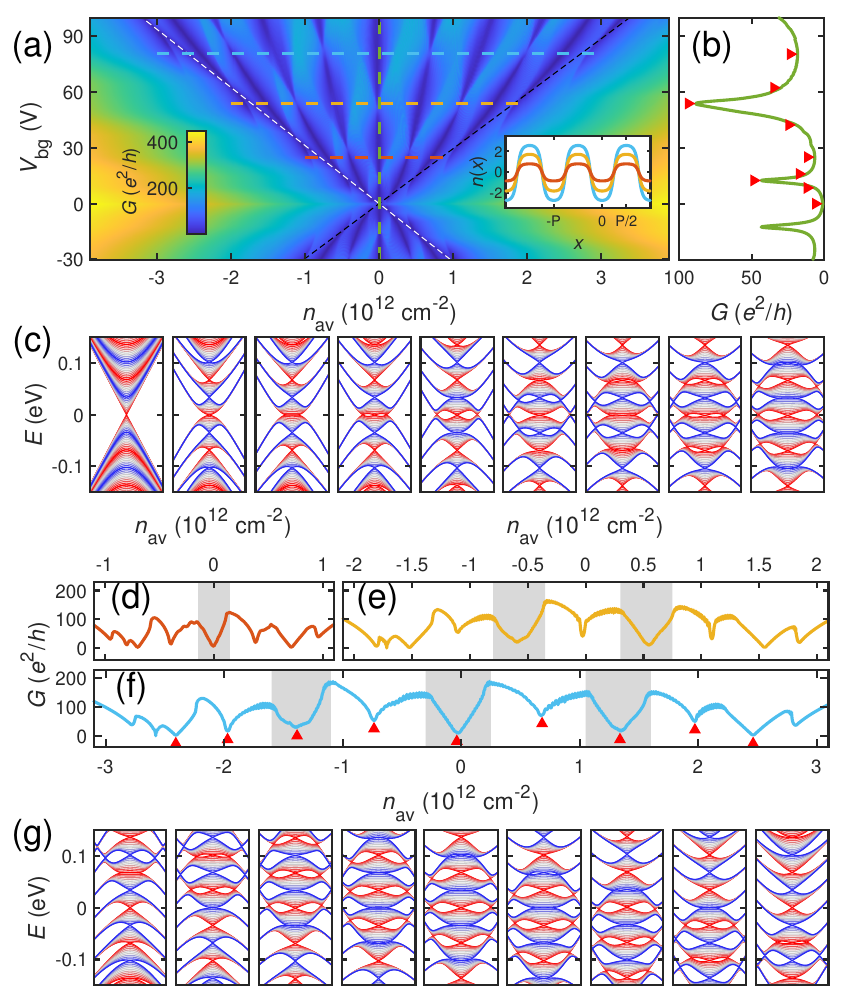}
\caption{Transport simulations and miniband structures of a device similar to \cite{Note1} but modified with $\ell=P/2=30\unit{nm}$. (a) Two-terminal conductance as a function of average density $n_\mathrm{av}$ and back gate voltage $V_\mathrm{bg}$. Diagonal black and white lines are defined in the main text. The vertical dashed line corresponds to (b) where $G(n_\mathrm{av}=0,V_\mathrm{bg})$ is shown, and the three horizontal dashed lines from bottom to top correspond to (d)--(f) where $G(n_\mathrm{av},V_\mathrm{bg}=25\unit{V})$, $G(n_\mathrm{av},V_\mathrm{bg}=53.75\unit{V})$, and $G(n_\mathrm{av},V_\mathrm{bg}=80.5\unit{V})$ are shown, respectively. Miniband structures in (c)/(g) correspond to red triangles marked in (b)/(g).}
\label{fig2}
\end{figure}

A small part of the conductance map based on the periodic model is overlayed on Fig.\ \ref{fig1}(b) marked by the white dashed box. The overall behavior is rather similar to the original map, but a moderate difference can be clearly observed. A line cut from this submap with the same gate voltage range is shown by the orange curve in Fig.\ \ref{fig1}(c), which can be directly compared to the density of states $D(E)$ at Fermi energy $E=0$ shown in Fig.\ \ref{fig1}(d). The consistency between the spectra of the conductance and Fermi density of states $D(0)$ is rather satisfactory, suggesting also the reliability of the following miniband structures. For graphene superlattices, conductance dips usually originate from the emerging Dirac points where the density of states significantly drops. Three exemplary miniband structures $E(k_y)$ are shown in Figs.\ \ref{fig1}(e)--(g), corresponding to the $n_\mathrm{tg}$ values marked on Fig.\ \ref{fig1}(c). Red and blue curves of $E(k_y)$ correspond to $k_x=0$ and $k_x=G_x/2$, respectively, as indicated by the color bar. Note that the same convention of coloring will be adopted in the miniband structures of Fig.\ \ref{fig2}. 

The periodicity $P$ along $x$ introduces the mini-Brillouin zone defined by the reciprocal lattice vector $G_x=2\pi/P$, so that $k_x\in [-G_x/2,G_x/2]$. On the other hand, $k_y$ is continuous because of the underlying continuum model for the miniband structure calculations \cite{Park2008a,Park2008b,Chen2019}. The chosen three examples show that the conductance dips can be categorized into at least three kinds: Dirac points at $k_x=\pm G_x/2$ [panel (e)], Dirac points at $k_x=0$ with additional zero-energy states [panel (f)] or without (see Fig.\ \ref{fig2}), and non-Dirac band edges at $k_x=\pm G_x/2$ between two neighboring $k_x=0$ Dirac points [panel (g)]. Figures \ref{fig1}(e)--(g) correspond to the increase of the average density $n_\mathrm{av}$ [see Fig.\ \ref{fig1}(c)], and hence the sinking of the bands due to the Fermi energy fixed at zero. This can be clearly seen by tracing the position of, for example, the $k_x=\pm G_x/2$ Dirac points (blue curves), which form structures similar to the Spider-Man mask. The Spider-Man-like feature arises from the $k_y=0$ Dirac point and the accompanying $k_y\neq 0$ Dirac points that are offset in energy, as a consequence of the asymmetric periodic potential \cite{Park2009a}. To generate zero-energy states as close in energy as possible, in the following we consider a device geometry similar to \cite{Note1} but with the etch size and periodicity modified to $\ell = P/2 = 30\unit{nm}$; see \cite{Note2} for the details.

\begin{figure*}[t]
\includegraphics[width=\textwidth]{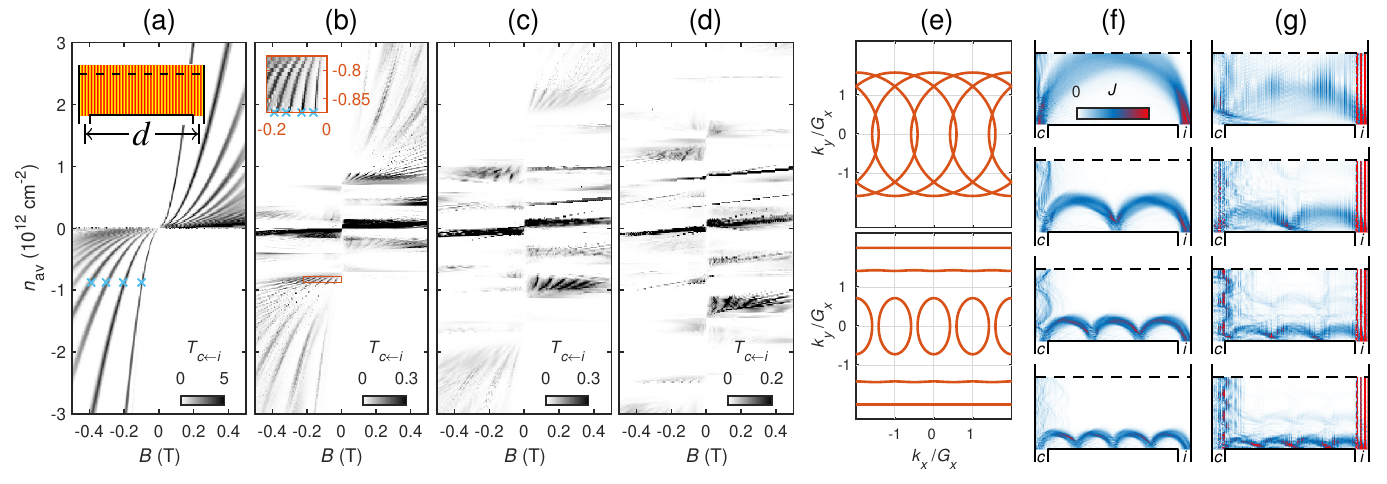}
\caption{Simulated transverse magnetic focusing on a three-terminal device [inset on (a)] with the electrostatic model considered in Fig.\ \ref{fig2}. Transmission function between injector and collector as a function of $B$ and $n_\mathrm{av}$ at (a) $V_\mathrm{bg}=0\unit{V}$, (b) $V_\mathrm{bg}=25\unit{V}$, (c) $V_\mathrm{bg}=53.75\unit{V}$, and (d) $V_\mathrm{bg}=80.5\unit{V}$. Blue crosses on (a) and (b) correspond to local current densities shown in (f) and (g), respectively. The Fermi contour for the unmodulated case of (a) and (f) is shown in the upper panel of (e), where the lower panel shows for the weakly modulated case of (b) and (g).}
\label{fig3}
\end{figure*}

To facilitate the following analysis, we define the composite gate voltage $V_g = (C_t/\langle C_b\rangle) V_\mathrm{tg} + V_\mathrm{bg}$, such that the average density is given by $\langle n\rangle \equiv n_\mathrm{av} = (C_t/e)V_\mathrm{tg}+(\langle C_b\rangle/e)V_\mathrm{bg} = (\langle C_b\rangle /e)V_g$. The simulated two-terminal conductance map $G(n_\mathrm{av},V_\mathrm{bg})$ is shown in Fig.\ \ref{fig2}(a), where the two diagonal dashed lines in white and black correspond to $n(x=P/2) = 0$ and $n(x=0)$, respectively. Contrary to the conductance map of Fig.\ \ref{fig1}(b) converted also to the $n_\mathrm{av}$-$V_\mathrm{bg}$ axes (see \cite{Note2}), the bipolar density range increases with the modulation strength of $V_\mathrm{bg}$ in a symmetric way. Moreover, the energy offset of the generated zero-energy states is much reduced, giving rise to a strong conductance suppression due to the aligned $k_y=0$ and $k_y\neq 0$ Dirac points, as illustrated below.

We first discuss the vertical line cut on Fig.\ \ref{fig2}(a) along $n_\mathrm{av}=0$, as shown in panel (b). The resonance peaks with the peak height increasing with the modulation strength controlled by $V_\mathrm{bg}$ resemble Fig.\ 1(a) of \cite{Brey2009}, where a model superlattice potential of $V(x) = U\cos(G_x x)$ is considered. It was found that the peaks correspond to the vanishing of the $y$-component group velocity $v_y = \partial E/\partial k_y=0$, where additional zero-energy Dirac points fulfilling $J_0(2U/\hbar v_F G_x)=0$, $J_n$ being the $n$th Bessel function of the first kind. At around $V_\mathrm{bg}=80\unit{V}$, the modulation amplitude estimated to be $U=V(P/2)-V(0)\approx 0.375\unit{eV}$ [see the inset of Fig.\ \ref{fig2}(a)] leads to $UP/4\pi\hbar v_F\approx 2.76$, suggesting two additional $k_y\neq 0$ zero-energy Dirac points, agreeing well with the line cut of Fig.\ \ref{fig2}(b), where the red triangles mark the gate configurations with which the miniband structures shown in Fig.\ \ref{fig2}(c) are calculated. 

From left to right panels of Fig.\ \ref{fig2}(c), the back gate voltage increases from the zero modulation of $V_\mathrm{bg}=0$ to strong modulation of $V_\mathrm{bg}=80.5\unit{V}$. The energy band structure evolves from the normal Dirac cone to the formation of minibands showing extra Dirac cones at $k_y=0$ and higher $E$, and eventually additional Dirac points at finite $k_y$ generated. Notably, at the rightmost panel of Fig.\ \ref{fig2}(c), the main $k_y=0$ Dirac point along with two $k_y\neq 0$ Dirac points are cloned at around $E\approx \pm 0.05 \unit{eV}$. As we will see below, these copies of Dirac points can also be accessed in transport by tuning the average density $n_\mathrm{av}$. 

We next focus on the three horizontal line cuts marked on Fig.\ \ref{fig2}(a) with weak (red), medium (orange), and strong (blue) modulations of $V_\mathrm{bg}=25\unit{V},53.75\unit{V},80.5\unit{V}$, shown in panel (d), (e), and (f), respectively. With the increasing modulation strength, the density range where the conductance exhibits multiple dips increases. As remarked previously, some of these dips arise from the main Dirac point accompanied by the generated zero energy states. By inspecting the miniband structures, we found that the number of such sets of Dirac points increases with the modulation strength: one set in Fig.\ \ref{fig2}(d), two sets in Fig.\ \ref{fig2}(e), and three sets in Fig.\ \ref{fig2}(f). Interestingly, these Dirac points appear at $k_x=0$ and $k_x=\pm G_x/2$ in an alternating pattern: an odd (even) number of such sets of Dirac points reside at $k_x = 0$ ($k_x=\pm G_x/2$). The density ranges in Figs.\ \ref{fig2}(d)--(f) corresponding to these sets of Dirac points are highlighted in gray. Figure \ref{fig2}(g) shows miniband structures corresponding to those conductance dips in Fig.\ \ref{fig2}(f) marked by the red triangles. Good consistency between Figs.\ \ref{fig2}(f) and (g) can be seen. Furthermore, the miniband structures of Fig.\ \ref{fig2}(g) show that the zero modes are not only cloned to higher energies but also accessible by tuning the average density $n_\mathrm{av}$.

We have further investigated transport properties of the modeled 1D graphene superlattice in the presence of a weak perpendicular magnetic field $B$. Instead of applying the B\"uttiker formula to calculate the four-point resistance \cite{Datta1995}, we consider a three-terminal geometry sketched in the inset of Fig.\ \ref{fig3}(a), and calculate the transmission function from the injector (right thin lead) to the collector (left thin lead), as a function of $n_\mathrm{av}$ and $B$, in order to minimize the computation burden. Considering various modulation strengths of $V_\mathrm{bg}$, the following analysis is meant to reveal gate-tunable ballistic miniband transport in the presently modeled 1D graphene superlattice. 

At zero modulation, our graphene sample is subject to uniform density $n_\mathrm{av}=n_\mathrm{tg}$, so that standard transverse magnetic focusing (TMF) \cite{Taychatanapat2013,Calado2014,Morikawa2015,Bhandari2016,Lee2016} is expected. As shown in Fig.\ \ref{fig3}(a), at least 5 TMF states can be seen, each satisfying the condition $d = j\cdot 2r_c, j = 1,2, \cdots$, where $r_c = \hbar \sqrt{\pi|n|}/eB$ is the cyclotron radius and $d=2200\unit{nm}$ is the probe spacing between the injector and collector leads. To be consistent with Figs.\ \ref{fig2}(d)--(f), we consider the same $V_\mathrm{bg}$ values in Figs.\ \ref{fig3}(b)--(d). With the increasing superlattice modulation, the formation of minibands can be seen, some of which exhibiting mini-TMF patterns. Contrary to the recent TMF experiment on the graphene/hBN moir\'e superlattice \cite{Lee2016}, here in Figs.\ \ref{fig3}(b)--(d), not only more minibands can be seen but also the formation of them can be tuned by gating. 

We have further imaged local current densities at $n_\mathrm{av}$ and $B$ values marked by the blue crosses on Fig.\ \ref{fig3}(a) for the zero modulation case and Fig.\ \ref{fig3}(b) for the weak modulation case. Their corresponding Fermi contours are shown in Fig.\ \ref{fig3}(e). Figures \ref{fig3}(f) and (g) reveal the first 4 TMF and mini-TMF states, respectively. Interestingly, in the weakly modulate case of Fig.\ \ref{fig3}(g), the elliptical cyclotron trajectories are observed, which arise from the simple geometric relation between the trajectories in reciprocal space and real space \cite{Ashcroft1976}.

\begin{figure}
\includegraphics[width=\columnwidth]{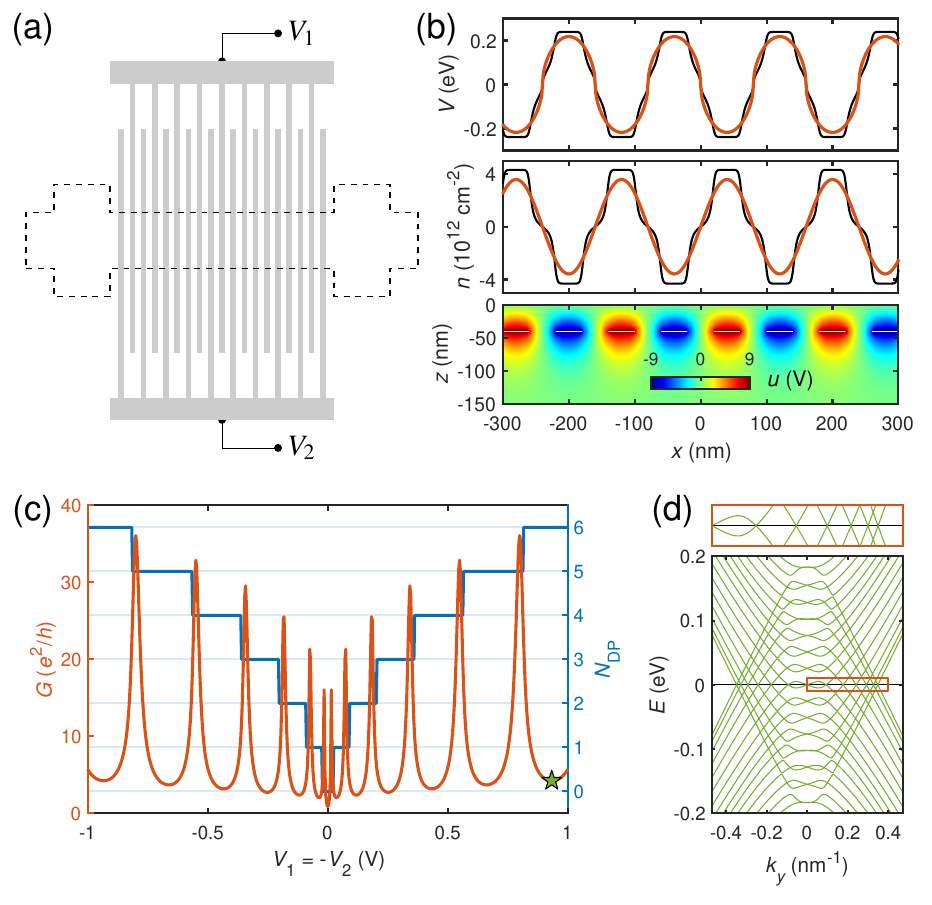}
\caption{1D graphene superlattice with strong and symmetric modulation. (a) Schematic of two sets of patterned few-layer graphene bottom gates, on top of which the graphene sample is to be placed (dashed Hall bar). (b) Exemplary electric potential distribution (bottom), carrier density profiles (middle), and onsite energy profile (top). (c) Simulated two-terminal conductance (red) and calculated number of Dirac points (blue) as a function of $V_1=-V_2$, assuming a 5-nm-thick hBN layer between patterned bottom gates and graphene. (d) Miniband structure corresponding to the star marked on (c); up to 6 additional zero modes are seen as shown on the top panel.}
\label{fig4}
\end{figure}

Finally, we propose a simple idea to achieve a perfectly symmetric periodic potential modulation for generating more zero-energy modes in 1D graphene superlattices, i.e., to achieve not only $V(x+P)=V(x)$ but also $V(x+P/2)=-V(x)$. From the electrostatic point of view, this cannot be achieved by using patterned substrate \cite{Note1,Forsythe2018}, but by using two sets of periodic local gates; see \cite{Note2} for a simple proof. We therefore consider another way of creating gate-controllable graphene superlattices using patterned few-layer graphene local bottom gates \cite{Drienovsky2017,Drienovsky2018}. Due to its stability and flexibility \cite{Drienovsky2017}, it shall be experimentally feasible to etch a few-layer graphene into two periodic local gates as schematically sketched in Fig.\ \ref{fig4}(a), where the two gates are controlled by gate voltages $V_1$ and $V_2$. Assuming a hBN/graphene/hBN stack transferred on top of these periodic gates, the resulting capacitance profiles can be rectangular-like or sine-like, depending on the thickness of the bottom hBN layer. %See \cite{Note2} for the simulated capacitance profiles, considering gate length and spacing both $40\unit{nm}$. 

The bottom panel of Fig.\ \ref{fig4}(b) shows the electric potential distribution around the periodic gates with $V_1=-V_2=9\unit{V}$ for the case of $40\unit{nm}$ thickness of the bottom hBN \cite{Note2}. The resulting carrier density and onsite energy profiles are shown by the red curves in the middle and top panels of Fig.\ \ref{fig4}(b), respectively, where the case of 5-nm-thick hBN is also shown (black curves). The perfectly symmetric periodic potential can be clearly seen and is expected to generate perfectly aligned zero modes. Figure \ref{fig4}(c) shows the two-terminal conductance as a function of $V_1=-V_2$ for the sharp case with 5-nm-thick bottom hBN, considering $10.5P$ of the scattering region ($P=4\times 40\unit{nm}$). Within $0<V_1 < 1\unit{V}$, up to 6 peaks can be clearly seen, indicating 6 additional zero modes generated, consistent with the calculated miniband structure shown in Fig.\ \ref{fig4}(d), which considers the gate voltages marked by the star in panel (c). We have also found that the previous formula prescribing the number of generated Dirac points (zero modes) \cite{Park2009a,Brey2009} can be slightly modified to $N_{PD} = \beta UP/4\pi\hbar v_F$ to match our result, where $\beta<1$ accounts for the considered realistic potential profile that is neither sinusoidal nor rectangular. With $\beta=0.71$, we found a perfect match as shown in Fig.\ \ref{fig4}(c).

In conclusion, we have shown how the signatures of 1D superlattices with zero modes generated would look like in standard transport measurements. In the presence of external magnetic field, we reveal mini-TMF patterns as signatures of ballistic miniband transport for the simulated 1D graphene superlattice. Our proposal for creating perfectly symmetric 1D superlattices that shall be experimentally feasible shows the possibility, within a reasonable gate voltage, of generating zero modes up to 6 which can be further accessed by tuning the average carrier density.

\begin{acknowledgments}

We thank P.\ Moon for stimulating discussions in DGIST and C.\ Dean for sharing their device geometry of \cite{Note1}. Part of the transport simulations in this work were performed using KWANT \cite{Groth2014}, whose authors are gratefully acknowledged. This work is financed by Minister of Science and Technology (MOST) of Taiwan under grant No.\ 107-2112-M-006-004-MY3.

\end{acknowledgments}

%%%%%%%%%%%%%%%%%% Bibliography %%%%%%%%%%%%%%%%%%
\bibliographystyle{apsrev4-1}
\bibliography{../../../../mhl2}

\newpage

\part*{Supplemental Material}

\renewcommand{\thefigure}{S\arabic{figure}}
\renewcommand{\theequation}{S\arabic{equation}}

%\appendix

\setcounter{figure}{0}

\section{Electrostatic models and simulations\label{supp sec 1}}

\subsection{Periodically etched substrate}

\begin{figure*}[t]
\centering
\subfigure[]{
\includegraphics[scale=0.5]{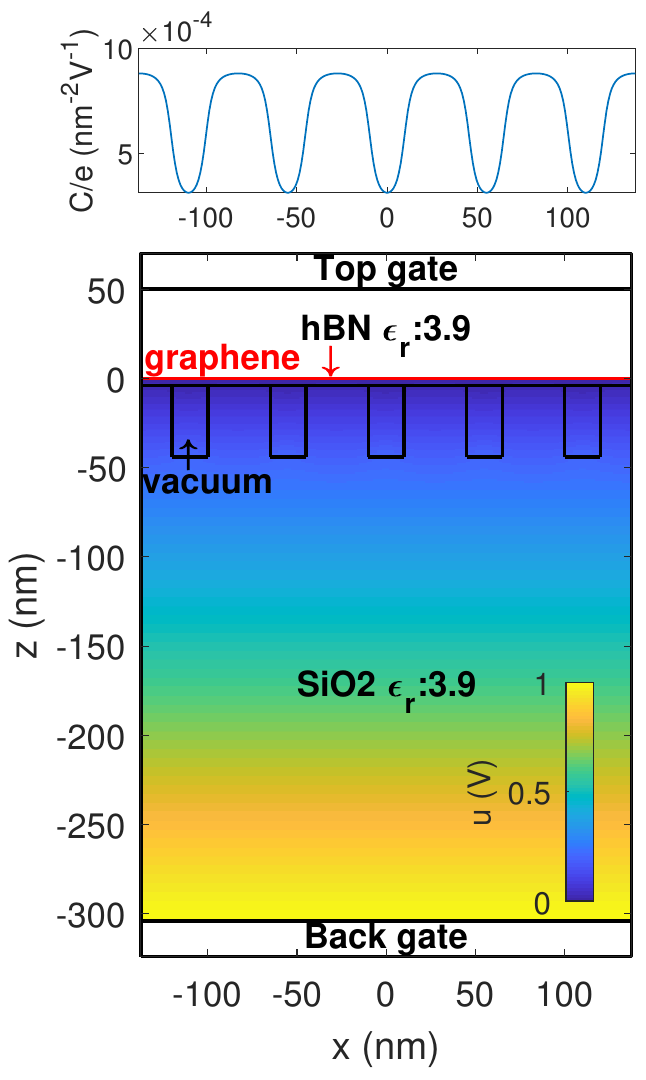}
}
\subfigure[]{
\includegraphics[scale=0.5]{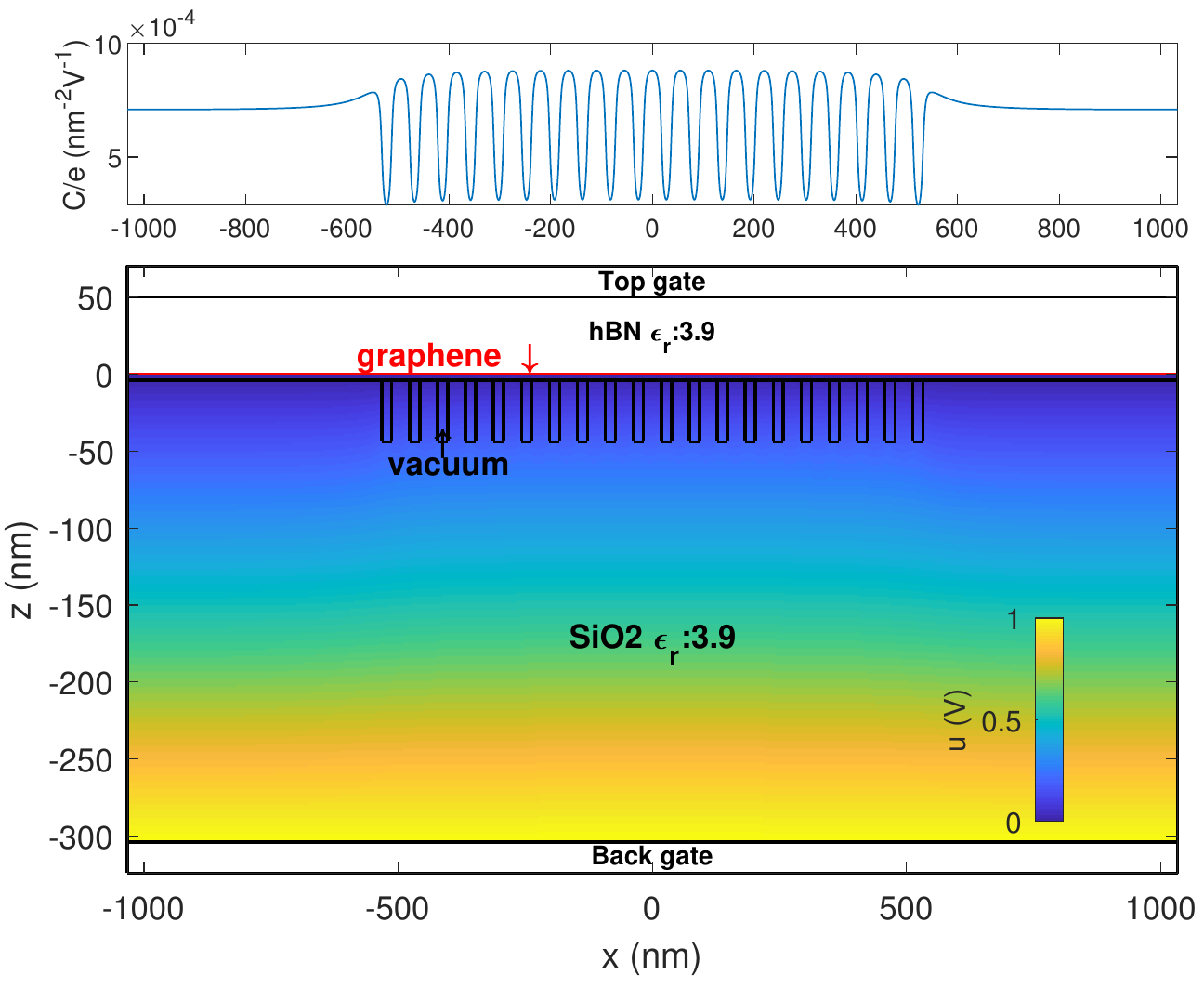}
}
\subfigure[]{
\includegraphics[scale=0.5]{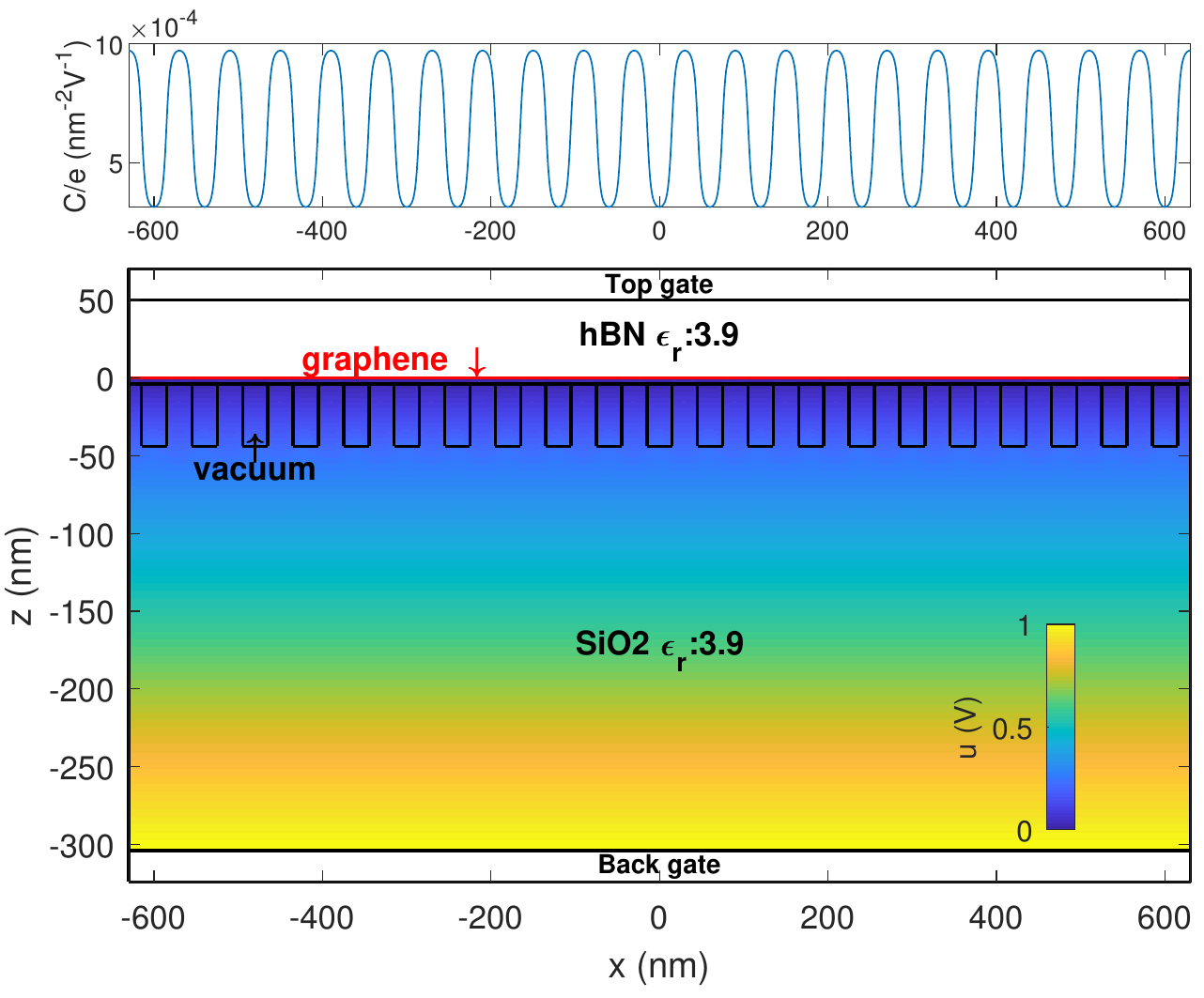}
}
\caption{(a) Periodic model with asymmetric etching. (b) Non-periodical model with asymmetric etching. (c) Periodic model with symmetric etching.}
\label{figS1}
\end{figure*}

The first three electrostatic models used in Figs.\ 1 and 2 of the main text are elaborated here. The modulated back gate capacitance is obtained by performing finite-element-based electrostatic simulation using \textsc{FEniCS} \cite{fenics} combined with the mesh generator \textsc{Gmsh} \cite{gmsh}. With \textsc{FEniCS}, the spatial distribution of the electric potential $u(x,z)$ is numerically solved, subject to given boundary conditions. At the boundary between graphene (set at $z=0$) and the bottom hBN layer, the surface charge density is given by
\begin{equation}
\sigma(x) = -\epsilon_r\epsilon_0\left.\frac{\partial u(x,z)}{\partial z}\right|_{z=0^-}\ ,
\end{equation}
where $\epsilon_r$ is the dielectric constant of hBN (set to 3.9 throughout this work) and $\epsilon_0$ is the free-space permittivity constant, and the spatially varying back gate capacitance (per unit area) is then defined as
\[
C_b(x) = \frac{\sigma(x)}{V_\mathrm{bg}}\ .
\]
Figure \ref{figS1} summarizes the three models used in Figs.\ 1 and 2 of the main text. The colorful regions show the electric potential distribution, and the resulting capacitance profiles are placed on the top.

Figures \ref{figS1}(a)--(c) consider similar parameters based on the device of \cite{Note1}: 4-nm-thick hBN between the 300-nm thick SiO$_{2}$ substrate and the graphene sample. A top hBN of thickness 50~nm is assumed but this is not needed for the electrostatic simulation. Both dielectric constants of hBN and SiO$_{2}$ are set to 3.9. The periodicity of the etching on SiO$_2$ is $P=55\unit{nm}$ in Figs.\ \ref{figS1}(a) and (b), and $P=60\unit{nm}$ in Fig.\ \ref{figS1}(c). The etch size is $\ell = 20\unit{nm}$ in Figs.\ \ref{figS1}(a) and (b), and $\ell=30\unit{nm}$ in Fig.\ \ref{figS1}(c). The structures of Figs.\ \ref{figS1}(a) and (c) generate periodic capacitance that was used for band structure calculations. On the other hand, Fig.\ \ref{figS1}(b) considers 21 trenches and a 500-nm-long buffer region at both ends, and is slightly closer to the experiment.

\subsection{Periodic few-layer graphene bottom gates}

\begin{figure*}
\subfigure[40-nm-thick bottom hBN]{
\begin{minipage}{0.48\textwidth}
\includegraphics[width=\textwidth]{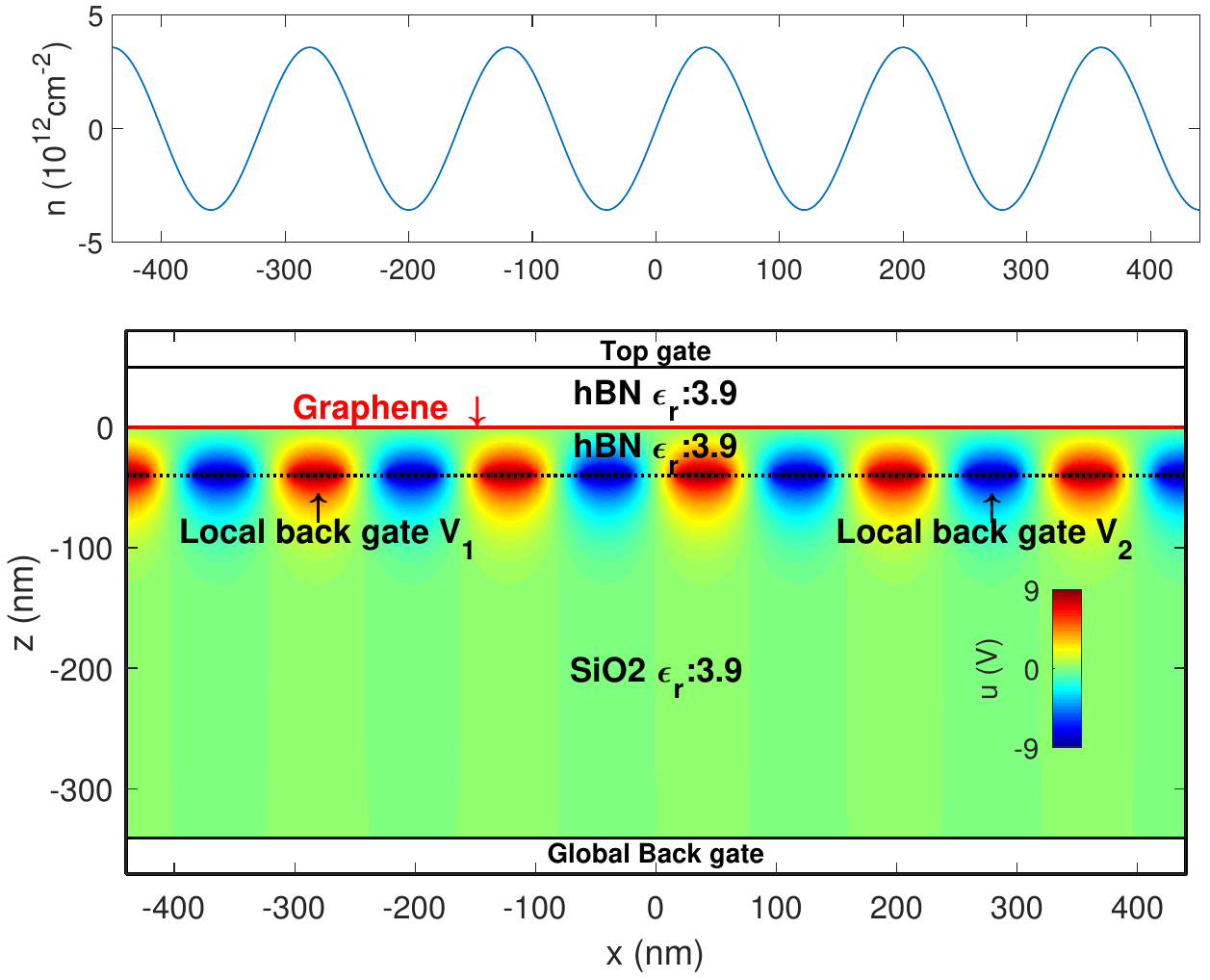}

\includegraphics[width=\textwidth]{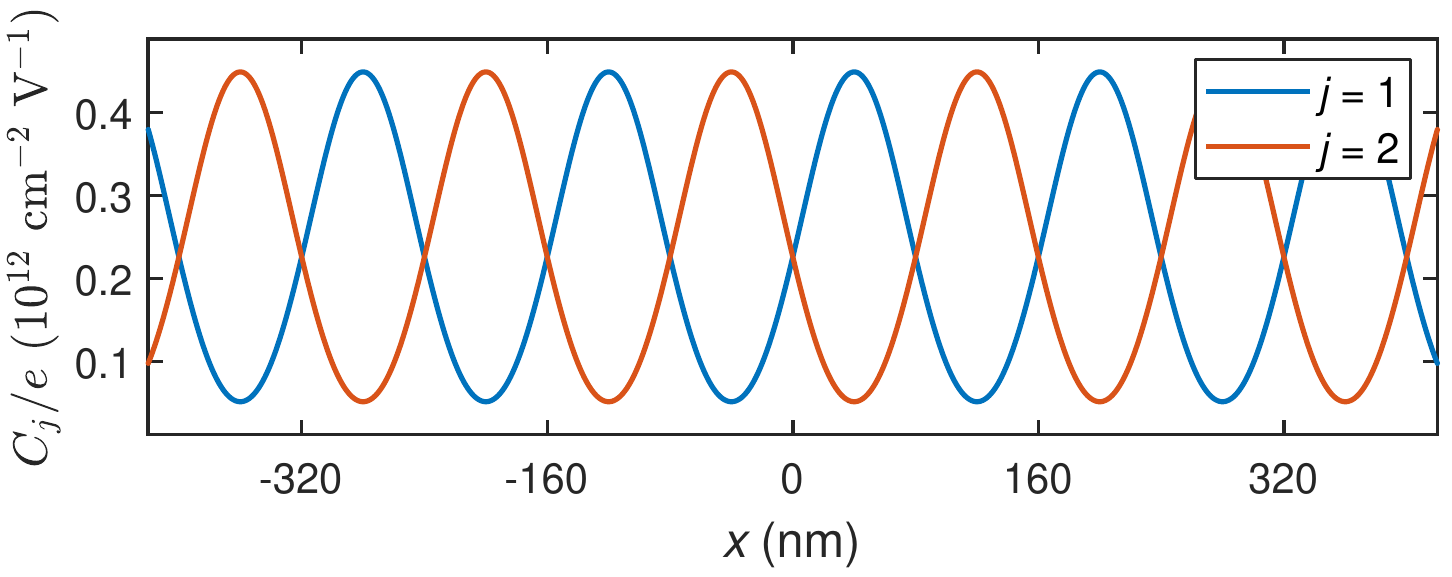}
\end{minipage}
}\hfill
\subfigure[5-nm-thick bottom hBN]{
\begin{minipage}{0.48\textwidth}
\includegraphics[width=\textwidth]{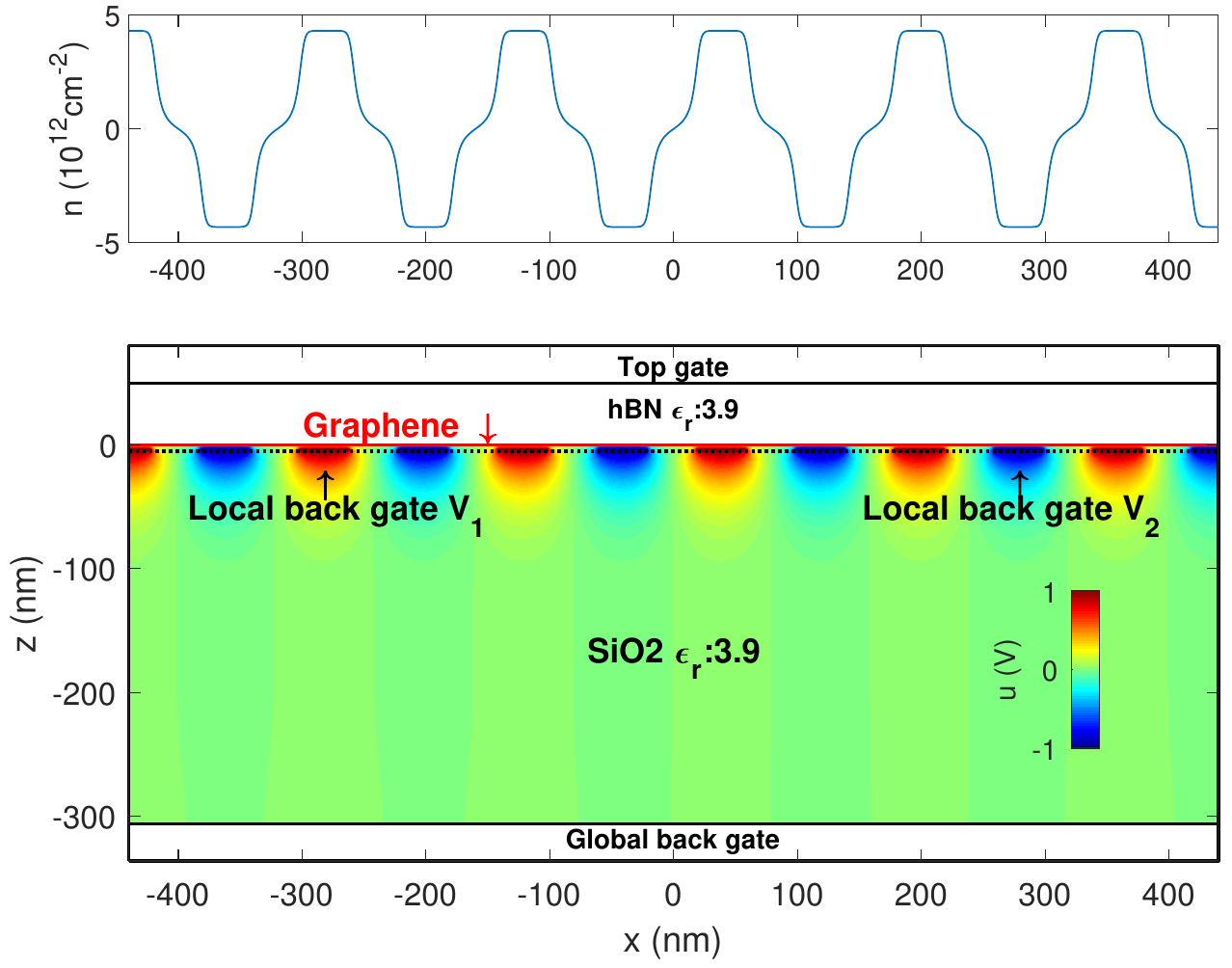}

\includegraphics[width=\textwidth]{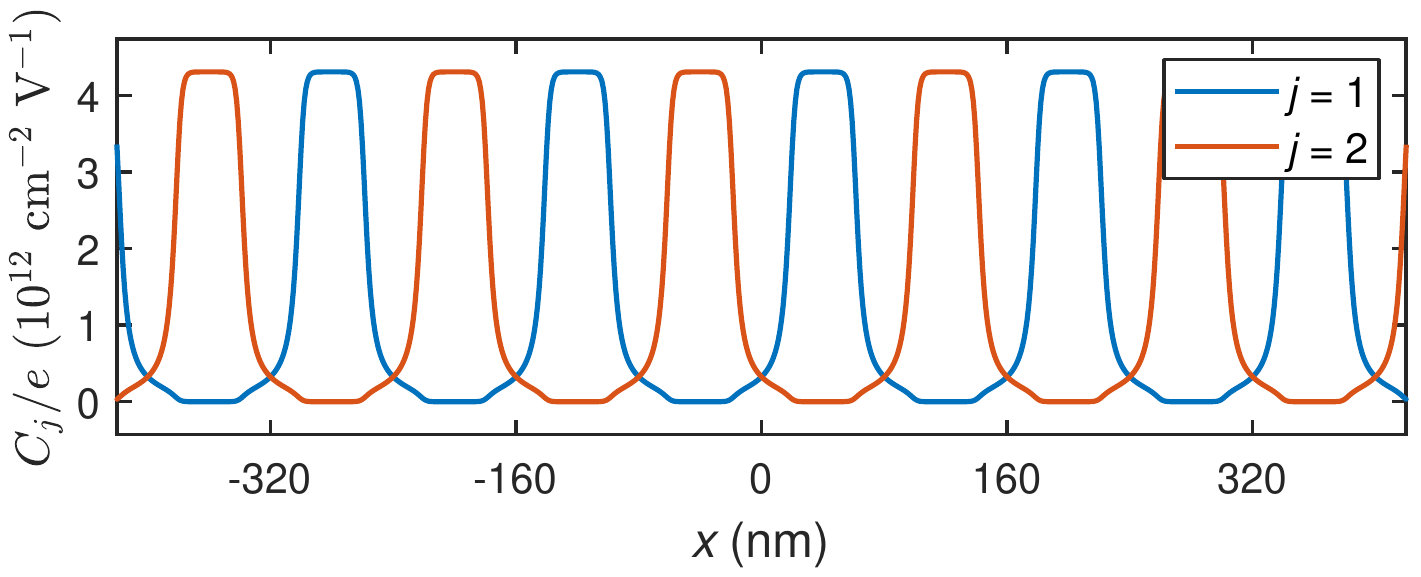}
\end{minipage}
}
\caption{Electrostatic models used in Fig.\ 4 of the main text.}
\label{figS2}
\end{figure*}

The proposed models in Fig.\ 4 of the main text use two sets of periodic local bottom gates to generate the perfectly symmetric periodic potential. These gates are assumed to alternate each other and lie on a 300-nm SiO$_{2}$ substrate followed by a global back gate. The graphene sample is assumed to be encapsulated by a 50-nm-thick top hBN layer and a bottom hBN of thickness 40 nm shown in Fig.\ \ref{figS2}(a) and 5 nm shown in Fig.\ \ref{figS2}(b). The length of all local bottom gates is 40 nm with 40 nm of spacing between them; the thickness of the few-layer graphene gates is set to 1 nm. The bottom panels in Fig.\ \ref{figS2} show the individual capacitance profiles $C_1$ and $C_2$; middle and top panels show exemplary $u(x,y)$ and $n(x)$, considering $V_1=-V_2=9\unit{V}$ and $V_1=-V_2=1\unit{V}$ in (a) and (b), respectively.

\section{More on simulated conductance maps\label{supp sec 2}}

\begin{figure*}
\includegraphics[width=\textwidth]{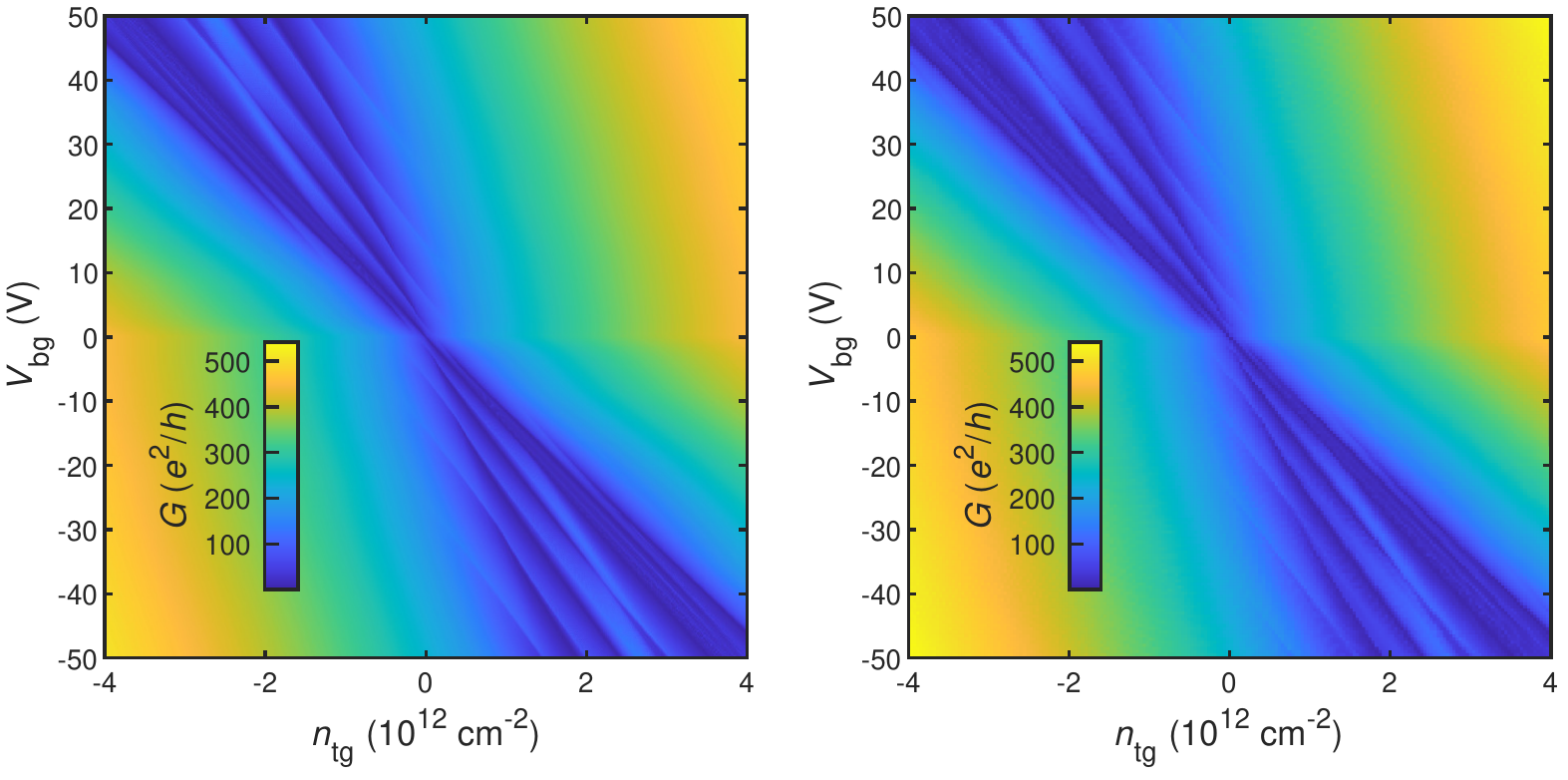}
\caption{Left: Conductance map simulated by $G=(W/3\pi a)g$ with $g$ obtained by using the $N_z=2$ zigzag graphene ribbon with periodic boundary hopping modulated by the Bloch phase. Right: Conductance map given by $G=(e^2/h)T$ with $T$ simulated by using an armchair graphene ribbon. In both cases, $W=1\unit{\mu m}$ is considered.}
\label{figS gVV vs GVV}
\end{figure*}

On Fig.\ 1(b) of the main text, we have overlayed part of the conductance map $G=(e^/h)T$ simulated by a $W=1\unit{\mu m}$ armchair graphene ribbon on top of the main conductance map simulated by $G=(W/3\pi a)g$ using periodic boundary hopping and showed stunning consistency. In Fig.\ \ref{figS gVV vs GVV}, the full maps are shown.

\begin{figure*}
\includegraphics[width=\textwidth]{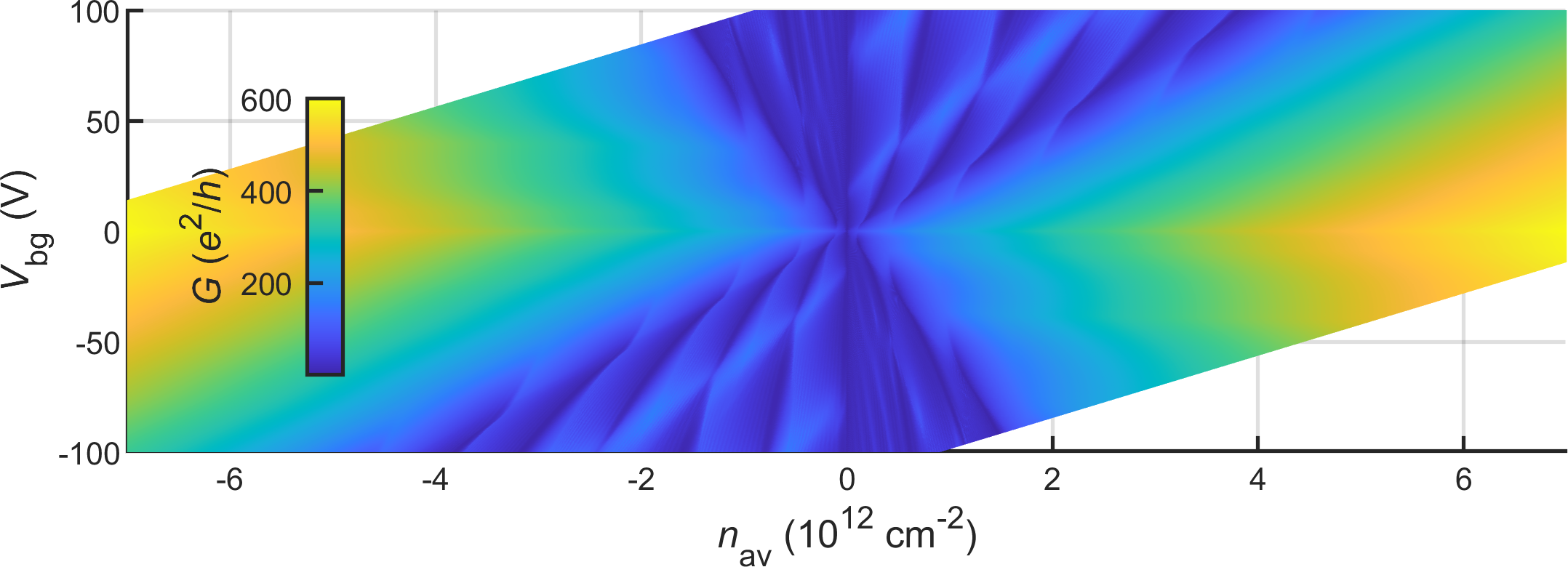}
\caption{Conductance map as a function of average density $n_\mathrm{av}$ and back gate voltage $V_\mathrm{bg}$ converted from the map of $G(n_\mathrm{tg},V_\mathrm{bg})$ shown in Fig.\ 1(b) of the main text.}
\label{figS GVV to GnV}
\end{figure*}

Figure \ref{figS GVV to GnV} shows the conductance map of Fig.\ 1(b) in the main text converted to the axes of average density and back gate voltage.

\section{Creating symmetric periodic potential\label{supp sec 3}}

In the main text, we have argued that the way of \cite{Note1} and \cite{Forsythe2018} using a periodically etched SiO$_2$ substrate cannot achieve a symmetric periodic potential. This is because the modulated back gate capacitance $C_b$ may fulfill $C_b(x+P)=C_b(x)$ but never the condition $C_b(x+P/2) = -C_b(x)$, such that $n(x)=[C_b(x)/e]V_\mathrm{bg}$ can fulfill $n(x+P/2)=-n(x)$ and hence $V(x+P/2)=-V(x)$. Using two periodic gates, this can be achieved, as we prove in the following.

Consider two periodic gate capacitances of periodicity $P$ that are offset to each other by a half period:
\begin{align}
C_1(x+P)&=C_1(x) \label{eq C1(x)} \\ 
C_2(x+P)&=C_2(x) \\
C_1(x+P/2) &= C_2(x) \\
C_2(x+P/2) &= C_1(x)
\end{align}
Apart from the uniform top gate that can be used to tune the average density, the carrier density profile due to these two sets of periodic gates is given by
\begin{align*}
n(x) &= \frac{C_1(x)}{e}V_1 + \frac{C_2(x)}{e}V_2\\ 
     &= \frac{C_1(x)}{e}V_1 + \frac{C_1(x+P/2)}{e}V_2\ .
\end{align*}
Now, considering symmetric bipolar gating, $V_1=-V_2 = V_g$, we have
\begin{equation}
n(x) = \frac{C_1(x)-C_1(x+P/2)}{e}V_1 = \frac{C(x)}{e}V_g\ ,
\label{eq n(x)}
\end{equation}
where we have defined a composite capacitance due to such gating as
\begin{equation}
C(x) = C_1(x)-C_1(x+P/2)\ .
\label{eq C(x) def}
\end{equation}
Because of Eq.\ \eqref{eq C1(x)}, we can use $C_1(x+P/2)=C_1(x-P/2)$ to rewrite Eq.\ \eqref{eq C(x) def} as
\begin{equation}
C(x) = C_1(x)-C_1(x-P/2)\ .
\label{eq C(x) alternative}
\end{equation}
The capacitance function $C(x)$ is another periodic function of period $P$ because of:
\begin{align*}
C(x+P) &= \underset{=C_1(x)}{\underbrace{C_1(x+P)}}-C_1(x-P/2+P)\\ &= C_1(x)-C_1(x+P/2) \overset{\eqref{eq C(x) def}}{=} C(x)\ ,
\end{align*}
and further fulfills
\begin{equation}
C(x+P/2) = -C(x)
\label{eq C(x+P/2) = -C(x)}
\end{equation}
because of
\begin{align*}
C(x+P/2) & \overset{\eqref{eq C(x) alternative}}{=} C_1(x+P/2)-C_1(x-P/2+P/2) \\ &= C_1(x+P/2)-C_1(x) \overset{\eqref{eq C(x) def}}{=} -C(x)\ .
\end{align*}
Equation \eqref{eq C(x+P/2) = -C(x)} suggests that the carrier density profile given by Eq.\ \eqref{eq n(x)} also fulfills
\[
n(x+P/2) = \frac{C(x+P/2)}{e}V_g = -\frac{C(x)}{e}V_g = -n(x),
\]
so that the resulting onsite energy profile $V(x)=-\hbar v_F\sqrt{\pi|n(x)|} \func{sgn}[n(x)]$ also fulfills $V(x+P/2)=-V(x)$.

A concrete numerical example assuming two sets of patterned periodic few-layer graphene bottom gate is given in \autoref{supp sec 1}.


\end{document}